\documentclass{elsart}
\usepackage{color}
\usepackage{graphicx}
\usepackage{epsfig}
\usepackage{amssymb}

\begin{document}
\begin{frontmatter}

\title{Production and trapping of radioactive atoms at the TRI$\mu$P facility}
\author{E. Traykov\corauthref{cor1}}{, }
\ead{traykov@kvi.nl}
\author{U. Dammalapati}{, }
\author{S. De}{, }
\author{O.C. Dermois}{,\,}
\author{L. Huisman}{,\,}
\author{K. Jungmann}{,\,}
\author{W. Kruithof}{,\,}
\author{A.J. Mol}{,\,}
\author{C.J.G. Onderwater}{,\,}
\author{A. Rogachevskiy}{,\,}
\author{M. da Silva e Silva}{,\,}
\author{M. Sohani}{,\,}
\author{O. Versolato}{,\,}
\author{L. Willmann}{,\,}
\author{H.W. Wilschut}
\address{Kernfysisch Versneller Instituut, University of
Groningen, Zernikelaan 25, 9747 AA Groningen, The Netherlands}
\corauth[cor1]{Corresponding author. Tel.: +31 503633569, fax: +31
503633401}

\begin{abstract}
The structures for the TRI$\mu$P facility have been completed and
commissioned. At the facility radioactive nuclides are produced
to study fundamental interactions and symmetries. An important
feature is the possibility to trap radioactive atoms in order to
obtain and hold a pure substrate-free sample for precision measurements. In the TRI$\mu$P facility a production target is followed by a magnetic separator, where radioactive isotopes are produced in inverse reaction kinematics. Separation up to 99.95$\%$ could be achieved for $^{21}$Na. A novel transmitting thermal ionizing device was developed to stop the energetic isotopes. Some 50$\%$ of stopped $^{21}$Na could be extracted and transported as low energy singly charged ions into a radio frequency quadrupole cooler and buncher with 35$\%$ transmission efficiency. The ions are transported lossless via a drift tube and a low energy electrostatic beam line into the experimental setup. Such ions can be neutralized on hot metal foils and the resulting atoms can be stored in a magneto-optical trap. The functioning of that principle was demonstrated with stable Na extracted from the thermal ionizer, radioactive beams will follow next.

\end{abstract}
\end{frontmatter}


\section{Introduction}
\label{intro}

Precision measurements of parameters describing $\beta$-decays and
searches for permanent electric dipole moments (EDMs) in atomic
systems are among the main objectives of the TRI$\mu$P physics
programme~\cite{wil99,
 jun02, wil03, jun05}.
Studies of $\beta$-$\nu$ correlations in nuclear $\beta$-decays
require the detection of very low energy recoiling ions. From these
correlations, deviations from the V-A structure of the weak
interaction can be searched for. Another line of research is to
improve the limits for permanent EDMs, which are
time-reversal violating moments and to improve the uncertainty of the values for parity non-conservation measuring the weak charge. For this research, advantage can be taken of special high sensitivity to new physics in certain heavy radioactive atoms~\cite{dzu01}.

The experiments will be performed with samples of radioactive atoms
stored in magneto-optical traps (MOT). Atomic trapping allows the storage of radioactive atoms without substrates and thereby boosts the performance of high precision experiments. Other advantages of atomic trapping include confinement and localization of the samples in space (typically in a volume $\leq$ 1 mm$^3$) at very low temperatures (in the sub-mK range), exclusive isotopic selectivity, and highly reduced background radioactivity.

At the Kernfysisch Versneller Instituut (KVI) a complex facility was
built both to achieve the goals of the TRI$\mu$P group and for
experiments with radioactive nuclides by external groups. The
facility employs the in-flight method for production and separation
of radioactive isotopes utilizing a dual-mode magnetic separator.
The separator is commissioned and operating~\cite{ber06}. Various
modes to produce radioactive particles have been tested for optimal
production~\cite{tra07b}. A thermal ionizer~\cite{tra07a} stops the
fast products and transports the nuclides as low-energy singly
charged ions into a radio frequency quadrupole (RFQ) cooler and
buncher~\cite{tra06}. This allows the collection and transport of
ions towards the optical traps. The optical trapping of Na atoms is
done in two separate MOT stages. The first MOT is used for
neutralization and accumulation of the Na atoms. It has been
commissioned with stable $^{23}$Na as described in~\cite{rog07}. The
atoms are then transported into a second MOT chamber where they can
be trapped again and their decays can be studied in a background-free environment~\cite{soh06}. The second MOT is equipped with $\beta$-detectors and a reaction microscope~\cite{kno01} for detecting of the recoils from the decay.

\section{Production and separation of radioactive nuclides}
\label{prod}

The TRI$\mu$P magnetic separator is used for in-flight production
and separation of a large variety of beam-like radioactive isotopes.
The nuclides are produced mainly using transfer reactions in inverse
kinematics (heavy projectile and light target) which provides a
secondary radioactive beam that can be matched to the angular and the momentum acceptance of the separator.
\begin{figure}
\centerline{\includegraphics[width=9cm,angle=0,clip=true]{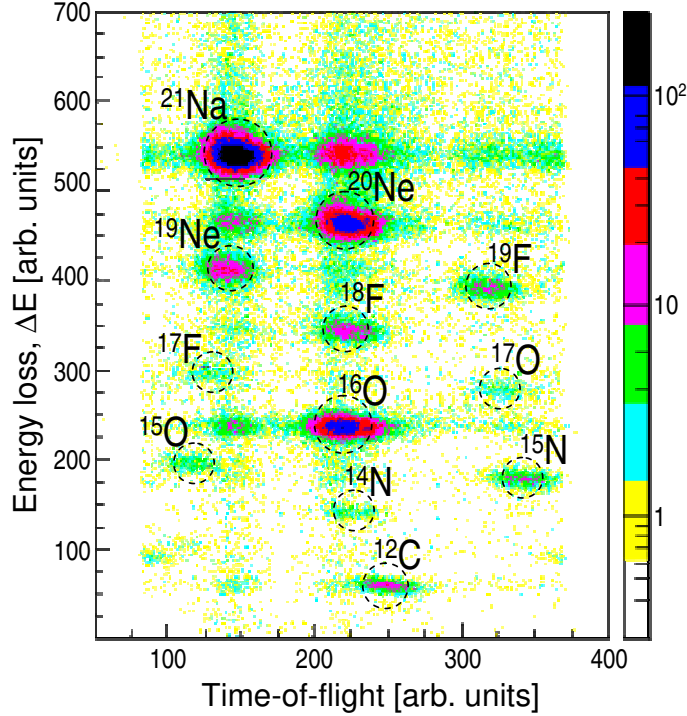}}
\caption {Energy loss vs. time-of-flight spectra measured with a
thin Si detector in the dispersive plane of the magnetic separator.
The plot is obtained for magnet settings optimized for the rate of
$^{21}$Na. When high purity $^{21}$Na is needed additional
purification is obtained using an achromatic degrader in the
dispersive plane of the separator. } \label{fig:21Na}
\end{figure}

The production target~\cite{you05} is filled with a light gas,
usually hydrogen, deuterium, or helium. The length of the gas volume
is 10 cm. The gas is kept at high pressure using Havar windows at
the two ends of the target. The windows are exchangeable and their
thickness can be varied depending on the gas pressure. Windows of
2.5 and 10 $\mu$m thickness allow safe operation at 1 and 10 atm
respectively. The target thickness is increased several times by
cooling of the target to liquid N$_2$ temperature.

The production yield of the desired isotopes in the target should be considered together with the selection in the separator when the aim is to obtain the highest isotope production rate at the end of the
separator. The factors contributing to the momentum distribution of
the secondary beam are related to the production mechanism, i.e. the
nuclear reactions, and to the angular and energy straggling in the
target and other materials in the separator. The interplay between
the cross sections for a specific reaction, the momentum and angular
distributions of the products, and the separator acceptances is considered for the choice of the initial energy of the primary beam.
%
For the production of $^{21}$Na various combinations of reactions
and beam energies were examined. We have found that the reaction
$^2$H($^{20}$Ne,$^{21}$Na)n is the most convenient, with a typical
yield of $10^4$/s/particle nA of primary beam at 1 atm. The
magnetic rigidity of $^{21}$Na is $6.4\%$ lower than the rigidity
of the $^{20}$Ne beam which allows their complete separation. When the separator settings are optimized for the rate of $^{21}$Na various other reaction products are also present in the spectra (see
Fig.~\ref{fig:21Na}). Additional purification (up to 99.95$\%$ for $^{21}$Na) is obtained by using an achromatic degrader in the dispersive plane of the separator. For the planned $\beta$-decay experiments the high yield of $^{21}$Na is of primary importance and degraders are not used. Mass and element selectivity is featured in the stages of the facility following the separator.

\begin{figure}
\centerline{\includegraphics[width=9.5cm,angle=0,clip=true]{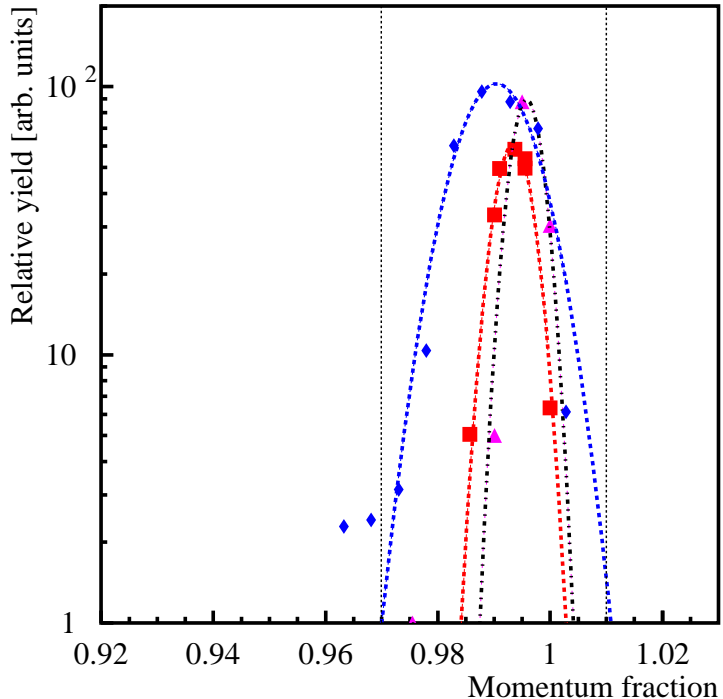}}
\caption {Production of $^{21}$Na: The momentum distribution
following the reaction $^2$H($^{20}$Ne,$^{21}$Na)n is shown for
three beam energies. A 10 cm long D$_2$  gas target is used cooled
to liquid N$_2$ temperature. Data labeled by {\tiny{$\blacksquare$}}
and $\blacktriangle$ were measured for 23 and 45 MeV per nucleon
beam energy respectively. 4 $\mu$m Havar windows and 1 atm D$_2$ gas
pressure were used. Data labeled by $\blacklozenge$ are for 10
$\mu$m Havar windows and 4 atm. The lines through the data points
are fitted gaussian distributions. The momentum acceptance
$\Delta{p}/{p}$ of the separator is 4\% (indicated by vertical
dashed lines).} \label{fig:mom}
\end{figure}

The $^{21}$Na momentum distributions are shown in more detail in
Fig.~\ref{fig:mom}. Extrapolating to higher pressures, we conclude
that a 10 atm target can be used, allowing the production of at least 10$^8$ $^{21}$Na particles/s. This will allow us to measure
$\beta$-$\nu$ correlations with a precision of 10$^{-4}$ in less
than one week of beam time. This is an order of magnitude
improvement over existing measurement~\cite{sev06}. It also
requires theoretical work to describe final state
interactions~\cite{vee02}.
%

Several experiments have been performed at the TRI$\mu$P facility
with the objective to measure properties of various radioactive
elements and their decays. Depending on the goals of the experiment, the optimization was focused on either element purity, or on maximal
production rate. Table 1 summarizes the isotopes produced to date
with the magnetic separator.

\begin{table}[ht]\label{Tab1}\caption{Produced radioactive isotopes; gas target at 1 atm.}
\centering
\renewcommand{\arraystretch}{1.0}
\vspace{\baselineskip}
\begin{tabular}{| c| c| c| c |c |c |c| }
  \hline
Product &   Beam    & Energy    &  Reaction & Target    &   Rate    &   Ref.    \\
        &           & [MeV/A]   &  type     &           & [/s/pnA]  &           \\
 \hline
 $^{21}$Na   &   $^{21}$Ne    &   20      &   (p,n)   &   CH$_2$     &                  &   \cite{ber06} \\
 $^{21}$Na   &   $^{24}$Mg    &   30      &   (p,a)   &   CH$_2$     &                  &                \\
 $^{21}$Na   &   $^{24}$Mg    &   30      &   fragm.  &   C          &                  &                \\
 $^{12}$B    &   $^{11}$B     &   22.3    &   (d,p)   &   D$_2$      &                  &   \cite{ped06} \\
 $^{12}$N    &   $^{12}$C     &   22.3    &   (p,n)   &   H$_2$      &                  &   \cite{ped06} \\
 $^{19}$Ne   &   $^{19}$F     &   10      &   (p,n)   &   H$_2$      & 1.1$\cdot$10$^3$ &   \cite{bro05} \\
 $^{20}$Na   &   $^{20}$Ne    &   22.3    &   (p,n)   &   H$_2$      & 1.0$\cdot$10$^4$ &   \cite{tra06} \\
 $^{21}$Na   &   $^{21}$Ne    &   43      &   (p,n)   &   H$_2$      & 3.0$\cdot$10$^3$ &   \cite{ach04} \\
 $^{21}$Na   &   $^{20}$Ne    &   22.3    &   (d,n)   &   D$_2$      & 1.3$\cdot$10$^4$ &   \cite{tra06} \\
 $^{21}$Na   &   $^{20}$Ne    &   40      &   (d,n)   &   D$_2$      & 8.0$\cdot$10$^3$ &   \cite{rog07} \\
 $^{22}$Mg   &   $^{23}$Na    &   31.5    &   (p,2n)  &   H$_2$      &                  &   \cite{ach04} \\
 $^{42}$Ti   &   $^{40}$Ca    &   45      &   (3He,n) &   $^3$He     &        20        &                \\
  \hline
\end{tabular}
\end{table}

\section{Slowing of radioactive ions}
\label{slow}

The achromatic focus of the separator coincides with the location of
a thermal ionizer (TI) which functions as an ion stopper. The
extraction efficiency from a TI can be close to 100\% for alkali
ions. The separator completely decouples the production site from the TI and the experimental setup. The stopping of the products in
the TI is accomplished using a stack of thin W foils which are
heated up to at least 2800 K. The total thickness of the foils is
chosen to match the maximum of the energy distribution
($\Delta{E}/E=\pm$ 4$\%$) of the ions at the focal plane of the
separator. The stopping range is adjusted by using a rotatable degrader upstream the TI. In the TI, only the ionized
particles are extracted but they have multiple chances to be ionized
inside the TI and to be electrostatically extracted by an electrode
at a negative potential (up to -10 kV). A yield increase from
the TI as function of temperature and of Na-isotope lifetime was
observed, but has not yet reached the point where the output is
saturated. A maximal efficiency of 48(3)$\%$ was achieved for
$^{21}$Na. The thermal ionizer principles, design, and operation are
described in more detail in~\cite{tra07a}.

Together with the radioactive Na isotopes, various stable ions are
present in the low energy beam extracted from the TI. These ions are
formed from diffusion of impurities in the W walls and foils of the
TI. In order to reduce space charge effects, mass selection is
desired. This is accomplished using a Wien filter downstream the
extractor electrode of the TI.

\section{Cooling and transport of low energy ion beams}
\label{cool}

\begin{figure}
\centerline{\includegraphics[width=13cm,angle=0,clip=true]{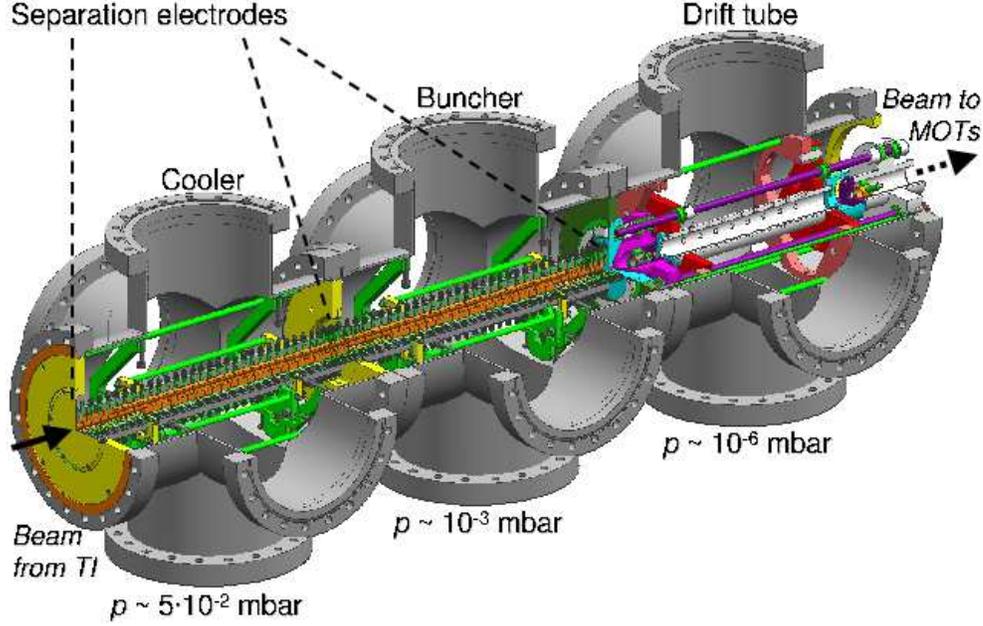}}
\caption {Technical drawing of the RFQ system (cooler, buncher, and
drift tube). The elements of the system are placed in standard NW160
vacuum chambers. Typical helium gas pressures are indicated.} \label{fig:rfq}
\end{figure}

The $^{21}$Na ion beam leaves the thermal ionizer with a large
transverse emittance and an energy distribution width defined by the
temperature of the W cavity. The large emittance is often affecting
the transmission of the ions in the beam line and, depending on the
requirements of the experimental setup, cooling may be required. This can be done by the radio frequency quadrupole (RFQ) cooler and
buncher system~\cite{tra06} as part of the low energy beam line (see
Fig.~\ref{fig:rfq}).

The cooling technique is based on collisions with a buffer gas in
combination with RF electric fields confining the ions in transverse
direction~\cite{deh67}. The buffer gas is composed of light atoms/molecules,
typically helium at pressures from $10^{-2}$ to $10^{-1}$ mbar. The RF electric field originating from 4 axisymmetric cylindrical
electrodes (330~mm long rods) creates a pseudo potential which
confines the ions between the rods. Opposite rods are separated
radially by $2r_0=10$~mm and potentials
$\pm\Phi=U_{DC}+U_{RF}cos(\omega_{RF}t)$ are applied to them.
To transport the ions through the cooler a DC drag potential is
applied in the longitudinal direction. It is obtained by dividing
the RFQ in longitudinal sections (36 segments per rod) allowing
different potentials to be applied. The DC potentials are set on the
separated sections whereas the RF potentials are applied on the
rods which are capacitively coupled with the segments. The
capacitative coupling is achieved using a 20 $\mu$m thin
Kapton$^{\textregistered}$ foil between the rods and the segments.

The cooled ions enter the buncher (mechanically identical to the
cooler) through a small aperture on an electrode plate separating
the cooler and the buncher and allowing differential pumping (Fig.~\ref{fig:rfq}). The
ion transmission through the aperture is critical because of the
damping of the RF fields near the aperture causing a reduction of
the radial confinement. To overcome this, a DC longitudinal
acceleration is applied before the aperture and a subsequent
deceleration in the beginning of the buncher.

The operation pressure in the buncher varies from $10^{-4}$ to
$10^{-2}$ mbar. This pressure allows the further reduction of the ion velocity and is sufficient to confine the ions in the
longitudinal direction by creating a DC potential well close to
the exit of the buncher. The trapped ions are then accumulated and
extracted in bunches by switching the DC potential configuration
when the trapping region is filled. The accumulation time is in
the range from several milliseconds to several seconds.

The ion bunches are accelerated to several keV in a drift tube
accelerator and guided into an all-electrostatic low energy beam
line where they are transported to the experimental sites. With the drift tube installed, both the RFQ and the low energy beam line can be kept on ground potential.

Monte Carlo simulations of the RFQ performance were made for Na ions
in order to estimate the effects of the fringe fields near the
apertures and to optimize the coupling of the RFQ sections and the
subsequent extraction. The interactions of the ions with the buffer
gas atoms were implemented in the simulations by taking into account
the diffusion related random force~\cite{rei65}. The amplitude of
the random force was normalized using data for ion mobilities in
gases.

The RFQ system was commissioned with stable Na ions. Various
measurements were made in order to characterize the system and
establish the operation ranges of various parameters, e.g.
potentials, pressures, accumulation times. Transmission was
optimized and efficiencies up to $60\%$ were measured for each of the RFQ stages. Measured storage times in the buncher showed strong
dependence of the buncher efficiency on the buffer gas purity and
the vacuum system.

The RFQ structures can also be exploited for mass selectivity. The variety of elements at a wide span of masses extracted from the
thermal ionizer was used to determine the mass selectivity of the
RFQ (Mathieu parameter $a\neq0$). Different ion masses were
identified using a microchannel plate detector after the drift tube
and measuring time-of-flight spectra of the ions. The optimal
selectivity was achieved by applying a potential $U_{DC}$ in the
buncher (due to lower operation pressures and longer trajectories
compared to the cooler). Selection was achieved for $\Delta{M}=2$ u
in the mass region of Na.

\section{Neutralization and trapping}
\label{neu}

\begin{figure}
\centerline{\includegraphics[width=12.6cm,angle=0,clip=true]{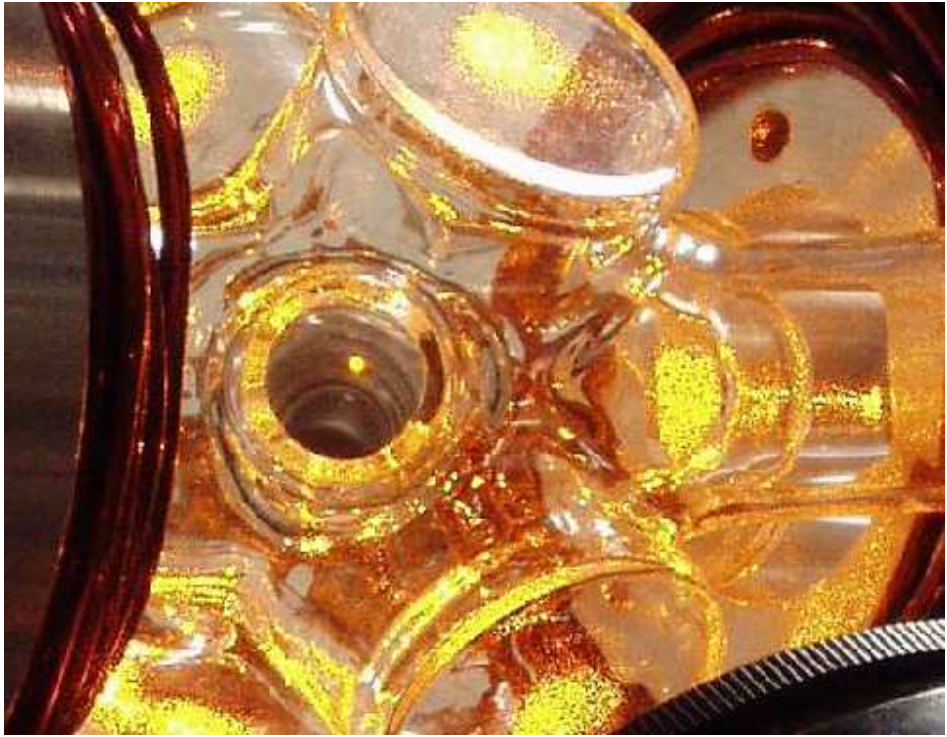}}
\caption {Photograph of an atomic $^{23}$Na cloud loaded into the
accumulator MOT. The fluorescence of the trapped atoms is visible in
the center of the glass cell.} \label{fig:mot}
\end{figure}

To trap atoms, the secondary ion beam needs to be neutralized. This
is done inside a small glass cell, the collection MOT~\cite{rog07} ,
using a heated neutralizer foil in which the ions are implanted with
energies up to a few keV. The neutralizer has thus common issues
with the the thermal ionizer, i.e. fast diffusion from a hot foil,
surface sticking time and ionization probability. In both cases
diffusion delay is the slowest time component in the process.
Various materials (e.g. Zr, Y, W, and LiF) were used for
neutralization of Na ions. Stable $^{23}$Na ions were used for
characterization of the neutralizers by observing the number of
captured particles in the MOT. Trapping was maximal for Zr which is
considered currently as the optimal neutralizer material.

\begin{figure}
\centerline{\includegraphics[width=10cm,angle=0,clip=true]{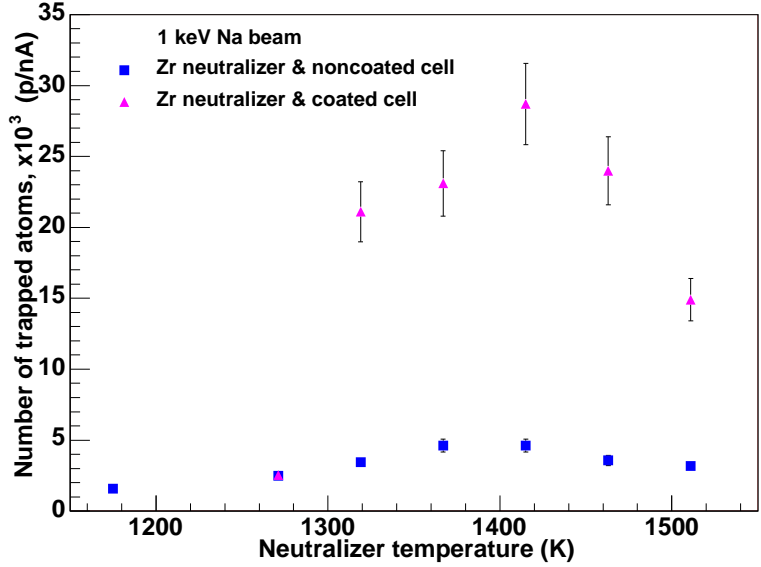}}
\caption {The number of trapped atoms in the MOT depends on the
neutralizer temperature. Optimal operation temperature is reached
when the effects of diffusion and vapor pressure balance. Dry film
coating of the glass cell increases several times the trapping
efficiency of the MOT for the same neutralizer material and
temperature range.} \label{fig:zr}
\end{figure}
Once neutral, the particles can be trapped in the MOT (shown in
Fig.~\ref{fig:mot}). The atoms in the cell have a thermal velocity
distribution (the neutralizer temperature) and only a fraction of
them are in the velocity range allowing trapping. The maximal
velocity for which atoms can be trapped can be increased by
increasing the red detuning of the lasers but require higher laser
power. The atoms which are not trapped when passing through the
trap region collide with the walls of the cell and may be lost due
to sticking and re-diffusion in the glass. The atoms desorbed from
the surface have lower velocities than the ones from the
neutralizer since their velocity distribution is defined by the
temperature of the walls.
Figure~\ref{fig:zr} shows two sets of data: from a glass cell without ({\tiny{$\blacksquare$}}) and with ($\blacktriangle$) a non-stick coating. The data obtained using the coated cell are higher in yield because the atoms can bounce of the walls and make multiple passes through the trapping region. The number of trapped atoms increases with temperature due to faster diffusion in the neutralizer. The fact that at higher temperatures the number of trapped atoms decreases
again is due to the increase of the pressure with temperature of the
neutralizer. The trap lifetime is inversely proportional to the
pressure.


The trapped atoms will be guided using lasers into a second MOT
which will be employed for the actual measurements of the
$\beta$-decay observables. In contrast to the first MOT cell, where
many atoms can decay without being trapped, in the decay MOT the
atoms will be well localized in the trap reducing the uncertainty of
the measurements.
This trap is situated in a large vacuum chamber including detectors
for the $\beta$ particles, a microchannel plate (MCP) detector with
position sensitive read-out, and guiding electrodes for the
detection of the recoiling $^{21}$Ne ions. The MOT setup for the
measurements of the $\beta$-$\nu$ correlations is currently being
commissioned~\cite{soh08}.

\section{Outlook}

A secondary beam of $^{21}$Na ions has been produced, purified, and
transferred successfully to a neutralizer in the glass cell around
the collection MOT site. We have exploited $^{20}$Na and $^{21}$Na
for the optimizing process using pairs of $\gamma$-ray detectors at
various places along the low energy beam line to observe the
annihilation radiation or $\beta$-delayed $\alpha$ decay
($^{20}$Na). Stable $^{23}$Na from the thermal ionizer is used as a
pilot beam in the low energy beam line. We are currently in the
process of optimizing the trapping and detection efficiencies in
preparation of our first physics experiment, a $\beta-\nu$
correlation measurement in the decay of $^{21}$Na. The successful
trapping and detection of $^{21}$Na in the MOT, which we expect very
soon as the next and remaining last step in a series of successful
preparation stages, will be the start of the physics program at the
TRI$\mu$P radioactive beam and trapping facility at KVI.


\end{document}